# Statistical and strict momentum conservation

Tian-Hai Zeng [1†], Zheng-Zhi Sun [1,2] and Bin Shao [1]

[1] *School of Physics, Beijing Institute of Technology, Beijing 100081, China*

[2] *School of Physical Sciences, University of Chinese Academy of Sciences, P. O. Box 4588, Beijing 100049, China*

**Abstract**

Arguments about the conservation laws of energy and momentum in the micro-world being statistical or strict began in 1924, and conflicting viewpoints remain today. The former is mainly supported theoretically, but the latter has been proved by many experiments. Here we explain that in principle, the strict conservation law of momentum always holds in the entangled state form of the momentum eigenstates of a closed composite system with interactions among subsystems, by expanding the total wave function to the sum of the products of momentum eigenstates of subsystems. Common scenario is that one of the two subsystems of a composite system is large or strong, its state remains approximately unchanged in a short time and the entangled state can be approximately written as a product state, which can be easily deduced from the paper of Haroche's group in 2001. The considered micro-subsystem can be approximately represented as the superposition of its different momentum eigenstates; therefore, the approximation can be used to explain why the law holds statistically for the subsystem and for any single particle due to neglecting the interactions with the large subsystem or the environment. So the two momentum conservation laws reasonably hold without conflicts.

**Keywords:** strict conservation law of momentum, entangled state**,** approximation, product state

**PACS:** 03.65.Ta

## 1. Introduction

In 1924, Bohr et al. [1] proposed the statistics-based conservation laws of energy and momentum ($P$) in the micro-world. Bothe and Geiger [2, 3] immediately tested the laws by doing experiments on the Compton effect, and their results showed that the strict laws hold in individual processes in 1925. Since then, arguments about the laws being statistical [4-8] or strict [9-31] have been presented, and the conflicting viewpoints have been given for nearly a century. Even today, no general consensus has been reached.

---

[†] Corresponding author. E-mail: zengtianhai@bit.edu.cn

The strict conservation law of a physical quantity in a system suggests that a measured quantity in the system or the addition of the same measured quantity to all subsystems is a constant quantity or a constant vector at any specific time (for example, before, after or in the process of the collision of two particles) in the system evolution process. If not, a conservation law is only of statistical meaning.

In 1926, Born formulated the famous statistical interpretation of the wave function, which was used to deduce that the laws in non-relativistic quantum mechanics are statistical, preventing the strict conservation laws from being included in quantum mechanics books [32-49], although the strict laws were validated by many experiments [10-15] after 1926.

The viewpoint of statistical laws in the micro-world was also supported by Dirac [5] and many authors in quantum mechanics books [32-49]. However, this perspective was mainly supported by theoretical formulations, and little support was based on experiments [4, 7, 8].

Considering different experimental forms of the Compton effect, the results of Shankland [4, 7] were statistical, but those of Bothe [2, 3] and Maier-Leibnitz [12], Hofstadter and McIntyre [13], and Cross and Ramsey [14], were strict. Other experiments [15] or explanations [16-18] of experiments supported the strict version of the laws. The neutron was discovered [19] and some experimental phenomena [20, 21] were explained using the strict laws. Some authors [22-31] have suggested that the conservation laws are strict and have applied the laws in theoretical analyses. Different from many books, only Zeng's books [37, 38] wrote about the strict conservation in the single processes.

Because interaction energy exists in the composite systems of all interacting subsystems and does not belong to any one subsystem, there is only a definition of the kinetic energy of a subsystem and not its energy. Under the condition that the interaction energy is approximated as the external potential energy for one subsystem, the energy can be considered the kinetic energy plus the external potential energy. For example, the external Coulomb potential energy of an electron in a free hydrogen atom is the approximation of the interaction energy between the electron and a proton. The measured energy of a subsystem should be its energy before or after interactions. However, the momentum $P$ of a subsystem can be measured at any given time. The law of the conservation of energy is more complex than that of $P$. Here we only discuss the conservation law of $P$.

2. **The momenta of single particles**

According to the principle of the superposition of states in non-relativistic quantum mechanics, which is used to describe systems with lower energy, the state of a free particle can be represented by the superposition of different $P$ eigenstates. If the momenta of identical particles in the same superposed state are measured separately, the results do not form a constant vector, and the conservation of $P$ is statistical for such singular particles.

In reality, any particle may not be free due to unavoidable interactions with its environment, so a free particle is only an ideal concept when the particle has a high energy and the interactions with the environment are neglected. The $P$ of a free particle is a continuous variable. Therefore, any tiny interaction between the particle and the environment can change the *P* of the particle and the *P* of the environment, which is the addition of the momenta of all the components of the environment. This change will occur because unavoidable interactions always exist. Thus, the conservation of *P* is statistical for any single particle, especially for particles with low energy in experiments.

Logically, it can be assumed that a free particle is in a *P* eigenstate or a superposed state of different *P* eigenstates because the states are the solutions of the Schrödinger equation for the free particle. When the *P* of a particle in a superposed state is measured, the particle is considered to be in a *P* eigenstate. However, the *P* eigenstate of a free particle cannot evolve into a superposed state of different *P* eigenstates according to the Schrödinger equation.

If a physical quantity *Q* non-commuting with *P* is measured for any state of a free particle, its state is considered as an eigenstate of *Q*. This eigenstate is theoretically considered to be a superposed state of different *P* eigenstates according to the principle of the superposition of states. But in the course of measurement, the particle must interact with a part of measuring equipment. In general, the composite system does not stay in a product state [33]. According to the Schrödinger equation, the system stays in an entangled state.

Recent experimental results [50-54] show that the entangled states have non-locality. From the non-locality, one may deduce that their entanglements still maintain even after the interactions between subsystems ceased and subsystems (particles) became free, and then each such free particle is in a mixed state.

If the interactions between subsystems cease, some possible results may be that the state becomes a product state of the *P* eigenstates of subsystems since no *P* exchanges between them, or becomes a product state of the *Q* eigenstates of subsystems after the *Q* is measured. But in experiment, a free particle in a *Q* eigenstate should interact with other object or strong field, its state then can be changed into an approximately superposed state of different *P* eigenstates, which will be explained in Sec. 4.

Therefore, we cannot assume that a free particle is in a superposed state (it is a pure state different from a mixed state) of different *P* eigenstates, and we can only assume that it is in a *P* eigenstate.

## 3. Entangled states and the law of conservation of momentum

The entangled state concept has been addressed in many papers and some books [38-49] over the past two decades, but it has not been used to discuss the conservation of *P*. Some recent papers [25-31] have discussed the law of the conservation of *P* without considering the concept of an entangled state.

We were able to find only one quantum mechanics book [55] in which the authors discussed the conservation law of *P* for a composite system of a photon and a beam splitter in an entangled state. In the entangled state, the magnitudes of the *P* changes of the photon and the beam splitter are equal, and the directions are opposite. However, the two states of the splitter, corresponding to two output photon states, in the entangled state are two different coherent states, which are superposed states of different *P* eigenstates. Thus, the addition of the measured momenta of the splitter and a photon do not form a constant vector, and this law is not yet strict.

A strict law can be obtained in a closed composite system with interactions among the subsystems if the interactions with the environment are weak and therefore neglected. The interactions change the momenta of the subsystems, but the change in the total $P$ is zero because the interactions with the environment are neglected.

The wave function of a free hydrogen (or any closed composite system of two subsystems with interactions between them) in a $\boldsymbol{p}$ eigenstate can be expressed as follows:

$$\Phi(\boldsymbol{r}_1, \boldsymbol{r}_2) = (2\pi\hbar)^{-3/2} \exp\left(\frac{i}{\hbar}\boldsymbol{p}\cdot\boldsymbol{R}\right)\psi(\boldsymbol{r}), \tag{1}$$

where $\boldsymbol{r}_1$ and $\boldsymbol{r}_2$ are the coordinates of the electron and the proton, respectively; $\boldsymbol{R} = (m_1\boldsymbol{r}_1 + m_2\boldsymbol{r}_2)/(m_1 + m_2)$ is the centre-of-mass coordinate; and $\boldsymbol{r} = \boldsymbol{r}_1 - \boldsymbol{r}_2$ is the relative coordinate. According to the principle of superposition of states, the wave function can be expanded with the addition of some terms that are the products of the $P$ eigenfunctions of the electron and the proton. Then, the wave function can be clearly expressed in the entangled state form of the $P$ eigenstates. Since the momenta have continuous spectra, the expansion has the following integral form:

$$\Phi(\boldsymbol{r}_1, \boldsymbol{r}_2) = (2\pi\hbar)^{-3} \int \phi(\boldsymbol{p}_1, \boldsymbol{p}_2) \exp(\frac{i}{\hbar}(\boldsymbol{p}_1\cdot\boldsymbol{r}_1 + \boldsymbol{p}_2\cdot\boldsymbol{r}_2))\, d\boldsymbol{p}_1 d\boldsymbol{p}_2. \tag{2}$$

The expansion coefficient is a $P$ space wave function:

$$\phi(\boldsymbol{p}_1, \boldsymbol{p}_2) = (2\pi\hbar)^{-3} \int \Phi(\boldsymbol{r}_1, \boldsymbol{r}_2) \exp(\frac{-i}{\hbar}(\boldsymbol{p}_1\cdot\boldsymbol{r}_1 + \boldsymbol{p}_2\cdot\boldsymbol{r}_2))\, d\boldsymbol{r}_1 d\boldsymbol{r}_2. \tag{3}$$

Substituting $d\boldsymbol{r}_1 d\boldsymbol{r}_2 = d\boldsymbol{r} d\boldsymbol{R}$ and Eq. (1) into Eq. (3) yields the following expression:

$$\phi(\boldsymbol{p}_1, \boldsymbol{p}_2) = (2\pi\hbar)^{-\frac{9}{2}} \int \exp\left(\frac{i}{\hbar}(\boldsymbol{p} - \boldsymbol{p}_1 - \boldsymbol{p}_2)\cdot\boldsymbol{R}\right) d\boldsymbol{R}$$

$$\times \int \psi(\boldsymbol{r}) \exp\left(\frac{i}{\hbar}(m_1\boldsymbol{p}_2 - m_2\boldsymbol{p}_1)\cdot\boldsymbol{r}/(m_1 + m_2)\right) d\boldsymbol{r}. \tag{4}$$

Therefore,

$$(2\pi\hbar)^{-3} \int \exp(\frac{i}{\hbar}(\boldsymbol{p} - \boldsymbol{p}_1 - \boldsymbol{p}_2)\cdot\boldsymbol{R})\, d\boldsymbol{R} = \delta(\boldsymbol{p} - \boldsymbol{p}_1 - \boldsymbol{p}_2). \tag{5}$$

The function $\delta(\boldsymbol{p} - \boldsymbol{p}_1 - \boldsymbol{p}_2)$ indicates that all non-zero expansion coefficients should fit $\boldsymbol{p} = \boldsymbol{p}_1 + \boldsymbol{p}_2$. Therefore, if the measured $P$ of the electron in a hydrogen is $\boldsymbol{p}_1$, the $P$ of the proton is definitely $\boldsymbol{p}_2 = \boldsymbol{p} - \boldsymbol{p}_1$, *i.e.*, the conservation of $P$ of the composite system is strict.

4. **Statistical conservation of momentum is approximate**

Haroche's group [56] described how particle *A* interacts with plate *B* and considered that the *A+B* system evolves into a combined or entangled state:

$$|\Psi\rangle = \frac{1}{\sqrt{2}}(|c\rangle|B_c\rangle + |d\rangle|B_d\rangle), \tag{6}$$

where $|c\rangle$ and $|d\rangle$ represent the particle wave packets in paths *c* and *d*, respectively, and

$|B_c\rangle$ and $|B_d\rangle$ represent the corresponding final states of $B$. They assumed two extreme cases: if $B$ is sufficiently light, $|\Psi\rangle$ is maximally entangled, and if $B$ is a heavy macroscopic object insensitive to the $P$ kick of a particle, $|B_c\rangle = |B_d\rangle$ and $|\Psi\rangle$ is an unentangled product state. However, the approximation $|B_c\rangle \approx |B_d\rangle$ may be a better description than $|B_c\rangle = |B_d\rangle$. They also wrote the combined atom + beam splitter (a coherent field) state, in a form similar to Eq. (6), as:

$$|\varphi\rangle = \tfrac{1}{\sqrt{2}}(|g\rangle|\beta_g\rangle + |e\rangle|\beta_e\rangle), \tag{7}$$

and the coherent field state $|\beta_g\rangle \approx |\beta_e\rangle$ in the classical limit, i.e., the mean photon number $N \gg 1$.

We can factor out $|B_c\rangle$ from Eq. (6) in the case of $|B_c\rangle \approx |B_d\rangle$:

$$\tfrac{1}{\sqrt{2}}(|c\rangle|B_c\rangle + |d\rangle|B_d\rangle) \approx \tfrac{1}{\sqrt{2}}(|c\rangle + |d\rangle)|B_c\rangle. \tag{8}$$

In this case, the entangled state can be approximately expressed as the product state of $|B_c\rangle$ and the superposition of $|c\rangle$ and $|d\rangle$, and this relation is similar [55, 57] for the entangled state (7).

The other result [56] of including the approximation in Eq. (8) is that the superposition of $|c\rangle$ and $|d\rangle$ for a particle is approximate and related with other objects or fields and the interactions among them.

Wineland's group [58] studied a laser beam interacting with an atom to produce the "Schrödinger cat" superposition state of the atom. In nature, this composite system would evolve into an entangled state of the laser beam and the atom, which can be approximately expressed by the product state of the approximately unchanged state of the laser beam and the "Schrödinger cat" superposition state of the atom. Thus, the state of the atom is related to the laser beam and to the interactions between the atom and the beam.

For simplicity, we may use $|p_c\rangle$ and $|p_d\rangle$ to express the $P$ eigenstates of the above particle in paths $c$ and $d$, respectively; $|P_c\rangle$ and $|P_d\rangle$ expressing the corresponding $P$ eigenstates of $B$. The state of the composite system is,

$$|\Psi\rangle = \tfrac{1}{\sqrt{2}}(|p_c\rangle|P_c\rangle + |p_d\rangle|P_d\rangle). \tag{9}$$

This form of entangled state can guarantee that the $P$ exchanged between the two subsystems is equal, $p_c - p_d = P_d - P_c$, or $p_c + P_c = p_d + P_d$, i.e., the conservation of $P$ of the composite system is strict. In the case of $B$ being a heavy macroscopic object, $|P_c\rangle \approx |P_d\rangle$, therefore,

$$\tfrac{1}{\sqrt{2}}(|p_c\rangle|P_c\rangle + |p_d\rangle|P_d\rangle) \approx \tfrac{1}{\sqrt{2}}(|p_c\rangle + |p_d\rangle)|P_d\rangle. \tag{10}$$

The right side of Eq. (10) is a product state, which means that the superposed state of different $P$ eigenstates of particle $A$ is approximately unrelated with plate $B$. But the left side of Eq. (10) means that $A$'s state is related with plate $B$. Therefore, the conservation of $P$ of particle $A$ is statistical. This is also the answer of the Sec. 2.

If we consider an environment, the entangled state of the three parts of $A$ and $B$ plus the environment guarantees that the addition of all the $P$ vectors yields a constant vector. Thus, such theoretical analysis results show that the strict conservation of *P* always holds in non-relativistic quantum mechanics, which have been supported by the above most experimental results.

5. **Conclusions**

Entanglements may guarantee the strict conservation of momentum in composite systems with interactions among subsystems. The approximations of entangled states become product states in many limit cases, and single particles or subsystems stay in approximate superposed states, which is why single particles or subsystems exhibit statistical conservation of momentum. Thus, this consideration can solve the conflict between statistical and strict conservation laws of momentum, and may aid in solving the corresponding conflicts for statistical and strict conservation laws of energy and angular momentum.

**Acknowledgment** Tianhai Zeng thanks Choo Hiap Oh, Berthold-Georg Englert, Gerard't Hooft, Christopher Monroe, Chinwen Chou, Tongcang Li, Chengzu Li, Linmei Liang, Guojun Jin, Shouyong Pei, Supen Kou, Xiaoming Liu, Haibo Wang, Shidong Liang, Zhibin Li, Keqiu Chen, Xianting Liang, Dianmin Tong, Sixia Yu, Guoyong Xiang, Yinghua Ji, Guiqin Li, Xuexi Yi, Yu Shi, Chun Liu, Fengli Yan, Hui Yan, Haosheng Zeng, Xinqi Li, Chengjie Zhang, Lifan Ying, Peizhu Ding, Xianguo Jiang, Liwen Hu, Chunyu Chang, Liyu Tian and my colleagues: Jian Zou, Xiusan Xing, Feng Wang, Xiangdong Zhang, Yugui Yao, Jungang Li, Hao Wei, Bingcong Gou, Yanxia Xing, Fan Yang, Zhaotan Jiang, Wenyong Su, Changhong Lu, Xinbing Song, Jianfeng He, Rui Wan, Haiyun Hu, Jinfang Cai, Yanquan Feng, Liang Wang, Rongyao Wang, Lin Li, Shaobo Zheng, Chengcheng Liu, Hongkang Zhao, Yongyou Zhang, Lijie Shi, Yulong Liu, Dazhi Xu, Lida Zhang, Shengli Zhang, Ye Cao, Jiangwei Shang, Anning Zhang, Fei Wang, and Hanchun Wu for their discussions and comments; Jinsong Miao for translations of early German literature; and Hao Wei, Guiqin Li, Lifan Ying, Bozhi Sun, Yongjun Lv, and Zhaobin Liu for their help. This work is supported by the National Natural Science Foundation of China under Grant Nos. 11674024 and 11875086.